\documentclass[apj,twocolumn,twocolappendix,floatfix]{openjournal}

\DeclareRobustCommand{\VAN}[3]{#2}
\let\VANthebibliography\thebibliography
\def\thebibliography{\DeclareRobustCommand{\VAN}[3]{##3}\VANthebibliography}

\usepackage{amsmath}	
\usepackage{amssymb}	
\usepackage{graphicx}	
\usepackage[breaklinks,colorlinks,citecolor=blue,urlcolor=blue,linkcolor=blue]{hyperref}
\usepackage{cleveref}
\usepackage[nobottomtitles]{titlesec}
\usepackage{natbib}
\titleformat{\section}{\filcenter\MakeUppercase}{\thesection.}{0.5em}{}

\usepackage{savesym}
\savesymbol{tablenum}
\restoresymbol{SIX}{tablenum}
\usepackage{needspace}
\usepackage{overpic}
\usepackage{xcolor}

\begin{document}
\title[V717 Andromedae]{V717 Andromedae: An active low mass ratio contact binary}


\author{\vspace{-1.3cm}S. S. Wadhwa$^{1}$, M. Grozdanović$^{2}$, N. F. H. Tothill$^{1}$, M. D. Filipovi{\' c}$^{1}$ and  A. Y. De Horta$^{1}$
}

\affiliation{$^1$Western Sydney University,\break Locked Bag 1797, Penrith, NSW 2751, Australia. 19899347@student.westernsydney.edu.au}
 \affiliation{$^2$Astronomical Observatory, Volgina 7, 11060 Belgrade 38, Serbia}

\begin{abstract}
Multi-band photometric analysis of the contact binary V717 Andromedae (V717 And) is presented. The system is found to be an extreme low mass ratio system ($q = 0.197$) with high inclination ($i = 84.5^{\circ}$), moderate degree of contact (27\%) and near equal temperatures of the components. Period analysis of survey photometry spanning over 6300 days reveals no significant change. Although the system does not demonstrate a significant O'Connell effect there are a number of other markers strongly suggesting the system is chromospherically active. Orbital analysis indicates that the system is stable and not a merger candidate.
\end{abstract}

\maketitle

\section{Introduction}


Close binary pairs where both components have expanded beyond their respective Roche lobes, resulting in bidirectional mass transfer and the formation of shared outer photospheric envelope, are referred to as contact binaries. Both components are still on the main sequence, but have expanded beyond their Roche lobes due to their close proximity and tidal effects. Although examples of high mass contact binaries are known, most examples have a primary (larger and heavier of the two components) with spectral class K to F with secondary component smaller and lighter than the primary. Regardless of the difference in the masses of the components, the temperature of the common envelope is close to that of the primary component. The secondary therefore is normally significantly brighter than its main sequence counterpart \citep{1981ApJ...245..650M, 2013MNRAS.430.2029Y}. Recently, there has been significant interest in the study of extreme low mass ratio contact binaries and theoretical considerations suggest that such systems may be unstable and potential merger candidates. Although quite a few extreme low mass ratio contact binaries have been found their predicted merger induced transient, a red nova, remains rare with only one confirmed event \citep{2011A&A...528A.114T}. The rarity of merger events suggests that pre-merger stage(s) is/are critical and poorly understood; identification of potential merger candidates therefore has gained considerable interest recently. Continued investigation of extreme low mass ratio contact binaries can potentially constrain the mechanisms underlying merger, fully define the instability mass ratio and potentially allow real time observation of stellar mergers.

The components and the common envelope are in a synchronous circular orbit, with current consensus suggesting long term evolution of contact binaries is driven by angular momentum loss (AML) secondary to magnetic stellar winds and magnetic breaking \citep{1981Ap&SS..78..401V, 2004MNRAS.355.1383L}. The components move closer together through continued AML, but the system can maintain synchronous orbit until the ratio of spin to orbital angular momentum approaches 3:1. The loss of synchronicity is thought to lead to rapid in-spiraling of the secondary component and potential merger of the components. During this phase, there is likely an exponential decrease in the orbital period. Although contact binaries are common, there is only one confirmed case of contact binary merger, that of V1309 Scorpii \citep[V1309 Sco,][]{2011A&A...528A.114T}. The combination of potential merger and increased magnetic activity has led to significant interest in both detection of magnetic activity through enhanced signatures of chromospheric activity \citep{2024ApJS..272....6H, 2024A&A...688A..23Z}, and orbital stability \citep{2021MNRAS.501..229W, 2024MNRAS.535.2494W}.

V717 And was identified as a contact binary system by \citet{2013PZP....13....9S}. We use the techniques described by \citet{2022JApA...43...94W} and various sky surveys to identify V717 And as a potential extreme low mass ratio contact binary system. The current study presents multi-band photometric analysis of V717 And. The geometric light curve solution combined with distance and extinction estimates are used to derive absolute parameters for the system. In addition, we used available spectra to look at markers of increased chromospheric activity, and finally, we evaluate the orbital stability of the system based on recent theoretical instability parameters for a system with similar mass primary.

\section{Target Selection and Photometric Observations}

Evolution modelling of contact binary systems has mostly suggested that the components will eventually merge following a period of angular momentum loss and possible non-conservative mass loss from the outer Roche lobe of the secondary or both components. Modelling has usually indicated merger taking place at extreme low mass ratios \citep{1977MNRAS.179..359R}. Theoretical considerations for orbital stability initially concentrated on the concept of a global minimum mass ratio at which merger would take place \citep{1995ApJ...444L..41R, 2007MNRAS.377.1635A}. More recently, however, this has been shown to be an over simplification, and it is now clear that each system has its own unique mass ratio at which it will merge and this mass ratio changes even during normal main sequence evolution \citep{2021MNRAS.501..229W, 2024MNRAS.535.2494W}. As part of our ongoing work in the identification of extreme low mass ratio, potentially unstable, contact binary systems, we have developed a rapid method of identification based on inspection of survey photometry \citep{2022JApA...43...94W}. The rapid identification routine indicated V717 And as a potential extreme low mass ratio system and we present an analysis of multi-band photometry along with a review of potential increased chromospheric activity, long-term period analysis, and of orbital stability.

V717 And was observed at the Astronomical Station Vidojevica, operated by the Astronomical Observatory of Belgrade, on September 1, 2024, using the $1.4\,\mathrm{m}$ ``Milanković" telescope equipped with an Andor iKon-L 936 CCD camera. 
Images were acquired using Johnson $B\,V\,R$ filters. Exposure times were 90 seconds, 45 seconds and 30 seconds for $B\,V\,R$ filters respectively. All images were calibrated using appropriate flat, dark and bias frames. Differential photometry was performed using the AstroImageJ software package \citep{2017AJ....153...77C} with 2MASS00122578+3124054 as the comparison star and 2MASS00122452+3122164 as the check star. Reference magnitudes of the comparison star ($V=14.37$, $B=15.51$, $R=14.04$) were adopted from \citet{2015AAS...22533616H}. The mean reported error for the estimated magnitudes was less than 0.005 mag for all filters. Basic observations of the light curves, summarised in Table 1, show a $V$-band amplitude of 0.41 mag with essentially equal maxima and minima. The $R$-band amplitude is 0.40 mag, again with near equal mimima and maxima. The $B$-band amplitude is higher at 0.48 mag, with the second maxima brighter by 0.03 mag and the primary minima fainter by about 0.03 mag. Such variations in amplitude and variation in maxima and minima between different bands are not uncommon among contact binaries \citep{2022RAA....22c5024X, 2011AJ....142..117S}.
\begin{table}

\caption{Light curve observations at quarter phase points for V717 And}
\centering
   \begin{tabular}{|l|l|l|l|l|}
    \hline
     &&&&\\
        \hfil Phase&\hfil B mag&\hfil V mag&\hfil R mag&\hfil B-V\\
         &&&&\\
        \hline
         &&&&\\
        \hfil 0 &\hfil15.02&\hfil 14.32&\hfil 14.18&\hfil 0.70\\
         &&&&\\
        \hfil 0.25 &\hfil 14.57&\hfil 13.92&\hfil 13.78&\hfil 0.65 \\
         &&&&\\
        \hfil 0.5 &\hfil 14.97&\hfil 14.31&\hfil 14.17&\hfil 0.67\\
         &&&&\\
        \hfil 0.75 &\hfil 14.54&\hfil 13.91&\hfil 13.78&\hfil 0.63 \\
         &&&&\\
        \hline
        
    \end{tabular}

    \end{table}

\section{Period Study}
There are no dedicated historical observations for V717 And. Survey data are available from the Catalina Sky Survey \citep{2014ApJS..213....9D}, All Sky Automated Survey - Supernova \citep{2014ApJ...788...48S, 2018MNRAS.477.3145J}, and the Zwicky Transient Facility Catalog of Periodic Variable Stars \citep{2020ApJS..249...18C}. The system was also observed by the SuperWasp survey \citep{2021MNRAS.502.1299T}, but the system is at the fainter limit of its instruments resulting in considerable scatter such that SuperWasp data was not used. In total, about 800 observations spanning 6350 days were available for review. Period changes, especially when small, are most accurately determined from long-term high-cadence observations. In the absence of such observations, the traditional Observed-Computed (O-C) technique offers no meaningful analysis. The technique of using periodic orthogonal polynomials to fit low cadence observations combined with an analysis of variance statistic to evaluate the quality of fit is regarded as being highly suitable for analysing unevenly and widely separated observation. The methodology is fully described in \citep{1996ApJ...460L.107S} and is widely implimented in various software packages including "Peranso" \citep{2016AN....337..239P}. Briefly, the data sample is divided into smaller subsets (often overlapping) and the "best" period for each subset calculated. The technique was used by \citet{2011A&A...528A.114T} in the period analysis of survey photometry, often using very small sample subsets, of the only confirmed contact binary merger event. Using the Peranso software package we divided the available survey data into multiple overlapping subsets of approximately 100 observations and determined the best period in the small range (0.25d to 0.35d) near the current estimate of 0.317357d. We find no significant change in the period of the system over 17 or so years of available observations. The changes in period over the survey time are illustrated in Figure 1. Based on our observations and the available survey data, we refine the ephemeris as follows:\\

\noindent$HJD_{min} = \\
2460555.54728(\pm261) + 0.317357015(\pm235)E$.\\ 

\begin{figure}
	\includegraphics[width=\columnwidth]{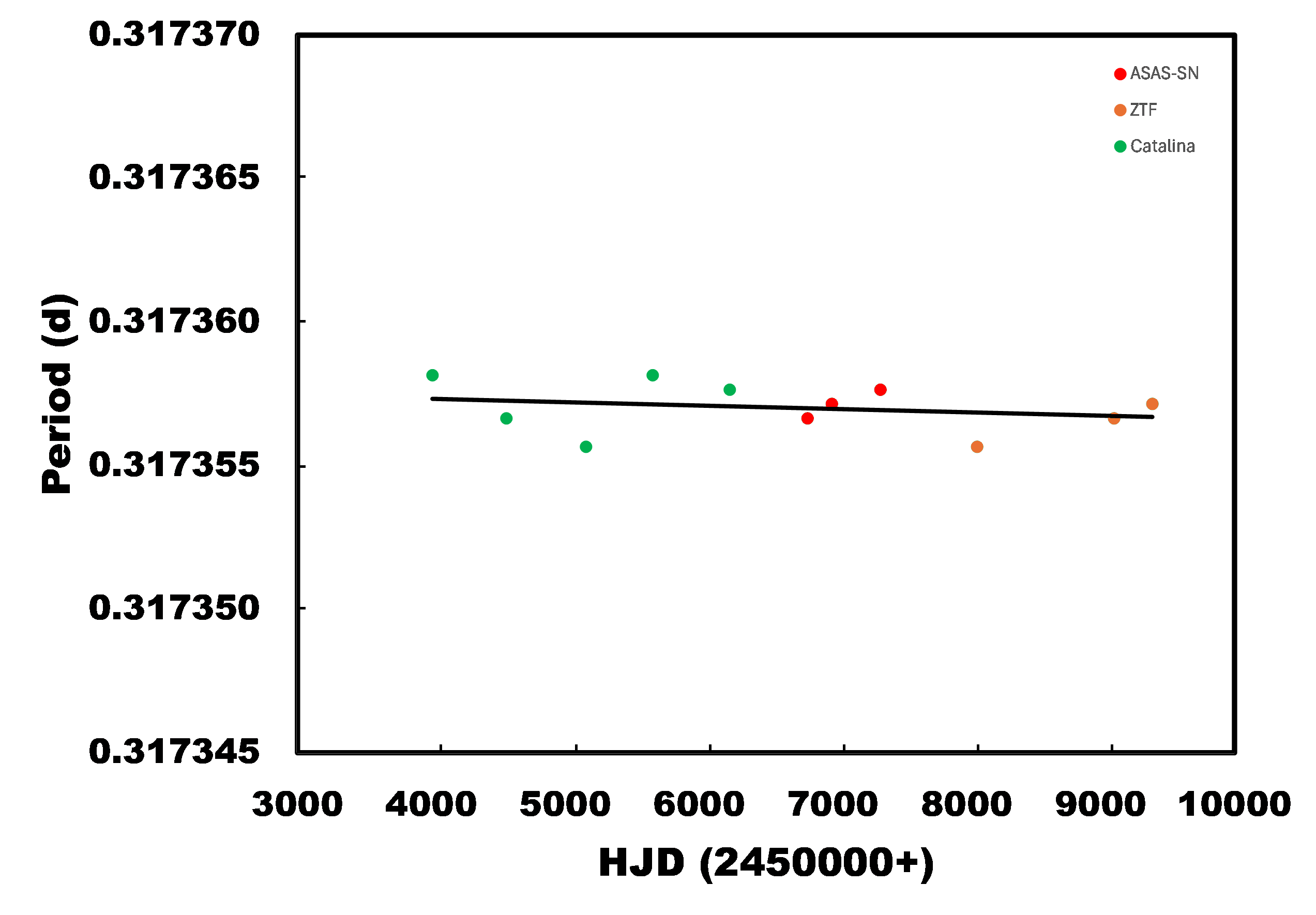}
    \caption{Period of V717 And spanning over 6300 days. Black line = Line of best fit. Estimates from various surveys as per colour coded data points.}
    \end{figure}

\section{Light Curve Analysis and Absolute Parameters}
The light curve of V717 And demonstrates complete eclipses and is therefore suitable for photometric determination of the mass ratio and geometric parameters \citep{2005Ap&SS.296..221T}. We used a recent version of the Wilson-Devinney (WD) code \citep{1971ApJ...166..605W, 1990ApJ...356..613W, 2021NewA...8601565N} to simultaneously model the $B\,V\,R$ photometric data. The widely used grid search method was used for values of the mass ratio ($q$) from 0.1 to 10 at intervals of 0.5 to find an approximate solution. Further refinement was performed around the approximate solution at intervals of 0.1 and finally at intervals of 0.01. During iterations the temperature of the secondary component ($T_2$), the potential (fillout) ($f$), inclination ($i$) and the dimensionless scaling factor, the luminosity of the primary ($L_1$), were the adjustable parameters. The fixed temperature of the primary ($T_1 = 5895K$) was adopted from the Large Sky Area Multi-Object Fiber Spectroscopic Telescope (LAMOST) spectra \citep{2022yCat.5156....0L}. As usual, the gravity coefficients were fixed at ($g_1 = g_2 = 0.32)$), bolometric albedoes ($A_1 = A_2 = 0.5$) were also fixed while the limb darkening coefficients were interpolated from \citet{1993AJ....106.2096V}. Simple reflection treatment was applied during analysis. As is standard practice, we adopt the standard deviation reported by the WD code on its final iteration as the errors for each estimated parameter. Potential random errors are not taken into account and as such the errors may be underestimated.

The system is found to be an extreme low mass ratio contact binary with a mass ratio $q=0.197\pm0.003$, high inclination $i=84.5^\circ\pm1.0^{\circ}$, moderate contact $f=27\pm7\%$ and near equal component temperatures, $T_2 = 5813\pm15K$. The geometric solution is summarised in Table 2. The critical portion of the grid search is illustrated in Figure 2. Observed and fitted light curves are illustrated in Figure 3. The variation in amplitude and the slight O'Connell effect noted in the $B$-band photometry is reflected in the slightly poorer fit of the observed and modelled light curves. We found no improvement in the fit with the inclusion of a third light.

\begin{figure}
	\includegraphics[width=\columnwidth]{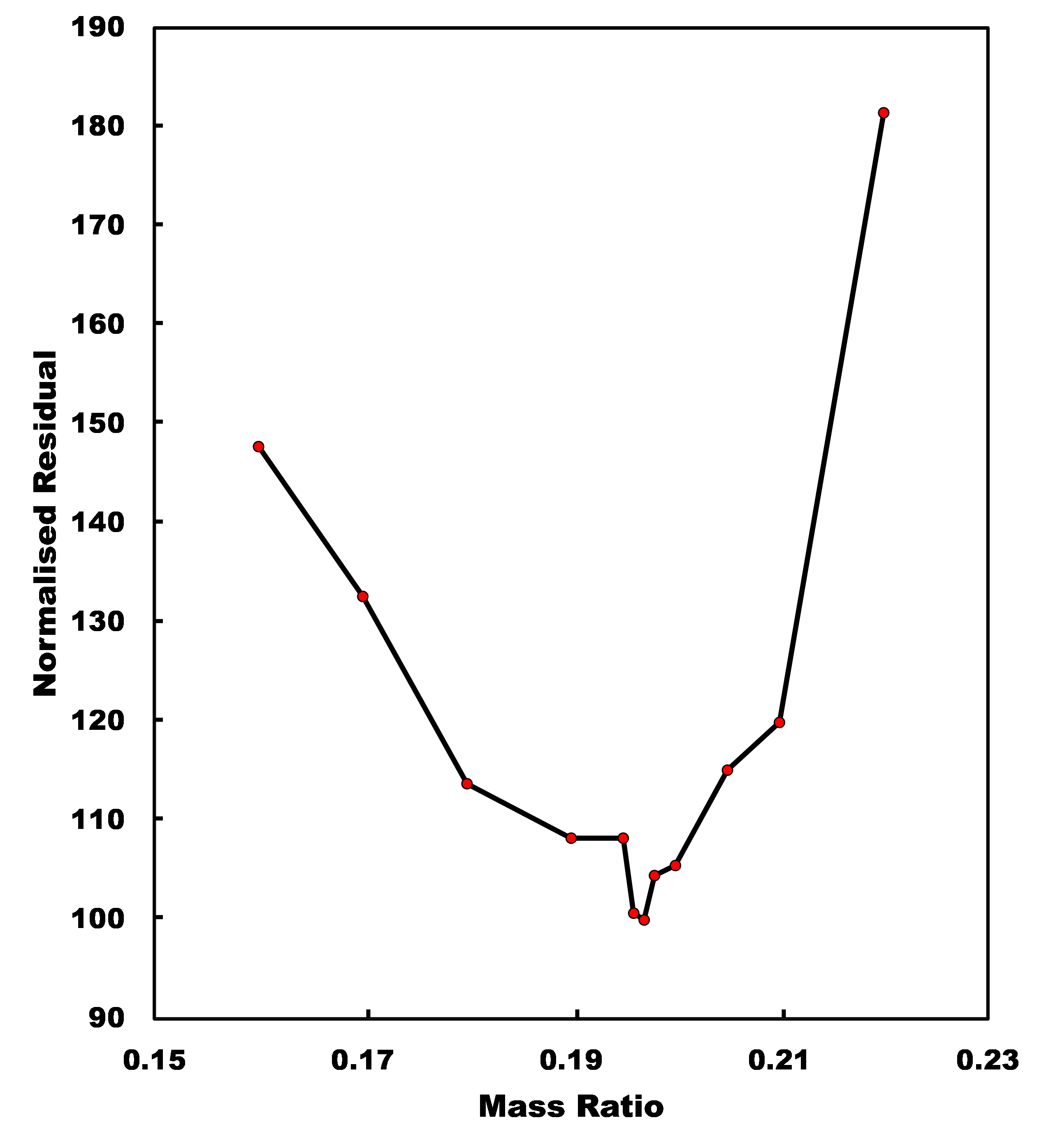}
    \caption{The critical portion of the mass ratio search grid from q=0.16 to q=0.22. The residual between the observed and fitted light curves has been normalised to the minimum residual at q=0.197}
    \end{figure}

As has been noted previously \citep{2021MNRAS.501..229W, 2022JApA...43...42W, 2020MNRAS.493.1565D}, analysis of survey photometric data can yield results in line with dedicated observations; however, there can be some variations, particularly with respect to the mass ratio and degree of contact. This is also true for V717 And for which analysis of survey photometry \citep{2024ApJS..271...32L} yields a mass ratio of 0.16, fillout of 0.43 and inclination of 80.4deg.

The most critical absolute parameter for further analysis is the mass of the primary component. In the absence of spectra, the luminosity of the primary component is usually used as an indirect method to estimate the mass of the primary component. The blackbody estimate of luminosity is based on the estimated radius and temperature of the primary component. Both of these parameters have some issues, noted below, which make blackbody estimate of luminosity not the ideal choice. As has been shown previously \citep{2022JApA...43...94W}, the radius of the primary component of contact binaries is significantly larger than that of the main sequence counterpart, and in the absence of formal spectra, colour based estimates of temperature can vary significantly and small variations in the assignment of the temperature can lead to significant variation in the estimate of the luminosity of a star if simple blackbody approximations are applied. For this reason, we prefer a more direct observational approach to estimate the mass of the primary component. As the light curve shows complete eclipses, we can conclude that the apparent magnitude at phase 0.5 is the apparent magnitude of the primary component. Using an accurate \emph{Gaia} \citep{2016A&A...595A...1G, 2023A&A...674A...1G} distance (862.4\,pc) and distance corrected extinction (0.162\,mag) as described in \citet{2023RAA....23k5001W}, we derive the absolute magnitude of the primary component, $M_{V1} = 4.47\pm0.04$. Using the April 2022 update \footnote{https://tinyurl.com/7yxyry8n} of \citet{2013ApJS..208....9P} tables we estimate the mass of the primary based on the absolute magnitude as $M_1 = 1.04M_{\odot}$, in good agreement with the spectral colour estimate from \emph{Gaia} of $0.97M_{\odot}$. Based on these two estimates, we adopt the mass of the primary as $M_1 = 1.00\pm0.03M_{\odot}$. All subsequent parameters and errors are propagated from this estimate. The mass of the secondary ($M_2 = 0.197\pm0.005M_{\odot}$) follows from the mass ratio, and the separation ($A=2.13\pm0.02R_{\odot}$) follows from the period and Kepler's third law. The radii of the components can be estimated as follows: ($R_{1,2} = A\times r_1$ and $A\times r_2$) where ($r_{1,2}$) are the geometric means of the fractional radii of the components in three orientations determined by the WD analysis \citep{2005JKAS...38...43A}. The absolute parameters are summarised in Table 2.

\begin{table}

\caption{Geometric and absolute parameters for V717 And. $r_{(1,2)}$ represent the geometric means of the fractional radii of the components. SMA = Semi Major Axis}
\centering
   \begin{tabular}{|l|l|l|l|}
    \hline
     
        \hfil Parameter&\hfil &\hfil Parameter&\hfil \\
        \hline
        &&&\\
         \hfil $T_1$ (Fixed) &\hfil5895K&\hfil $M_1$($M_{\odot}$)&\hfil1.00$\pm0.03$\\
        \hfil $T_2$ &\hfil 5813$\pm15$K&\hfil $M_2$($M_{\odot}$)&\hfil 0.197$\pm0.005$\\
        \hfil $i^{\circ}$ &\hfil 84.5$\pm1.0^{\circ}$&\hfil $R_1$($R_{\odot}$)&\hfil 1.14$\pm 0.03$ \\
        \hfil $q$ &\hfil 0.197$\pm0.003$&\hfil $R_2$($R_{\odot}$)&\hfil 0.56$\pm0.02$ \\
        \hfil Fillout ($f$)&\hfil 27$\pm7$\%&\hfil SMA ($R_{\odot}$)&\hfil 2.13$\pm0.03$\\
        \hfil $r_{1,2}$&\hfil 0.53525, 0.26142&\hfil &\hfil\\

        \hline
        
    \end{tabular}

    \end{table}

\begin{figure}
	\includegraphics[width=\columnwidth]{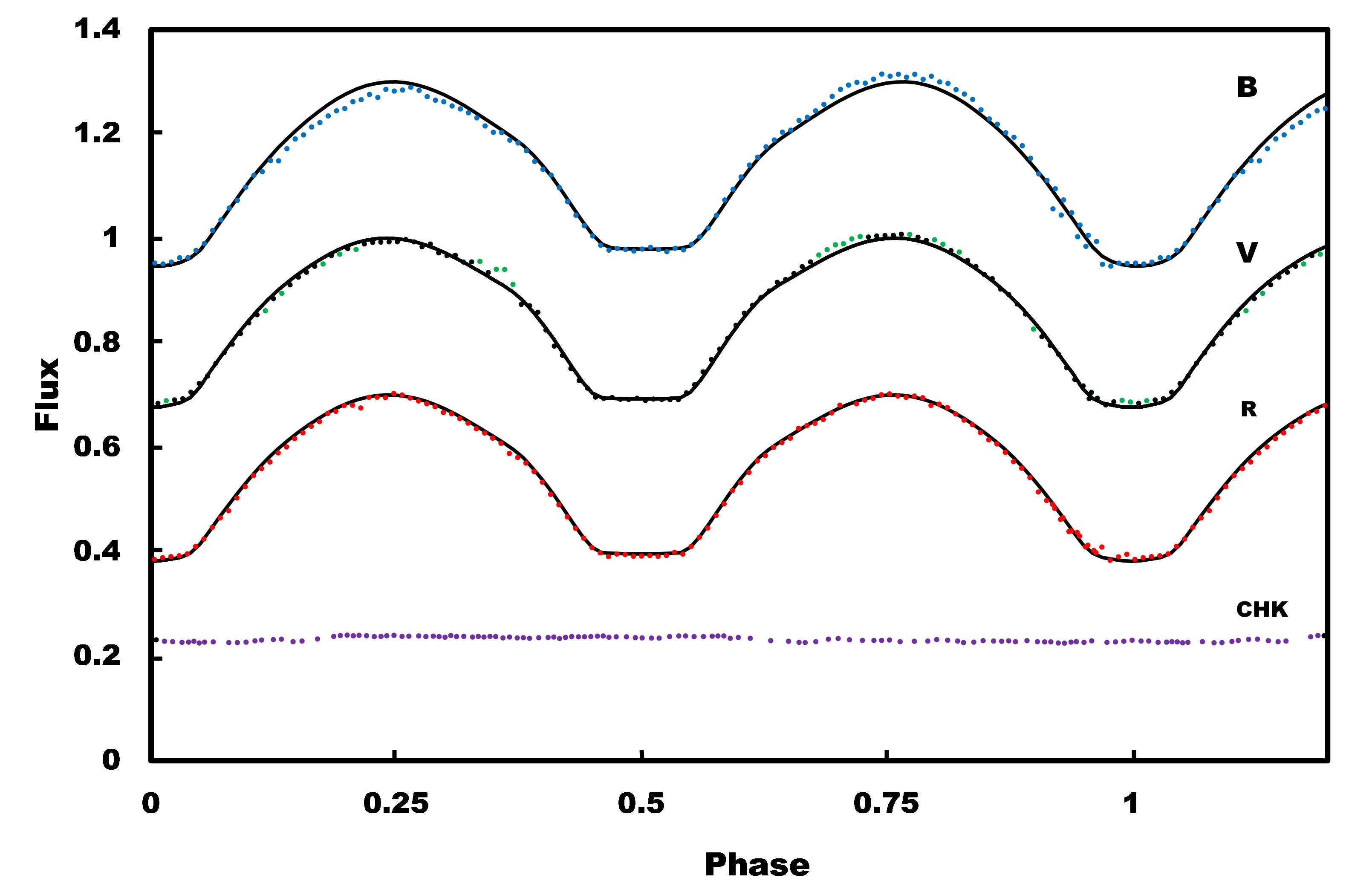}
    \caption{Observed and modelled light curves for V 717 And. The blue curve shows $B$-band, green, $V$-band, and red, $R$-band; the purple curve shows the Check Star. Black lines represent the WD modelled light curves for each band. The curves have been shifted vertically for clarity.}
    \end{figure}

\section{Chromospheric Activity and Orbital Stability}
\subsection{Chromospheric Activity}

Markers of enhanced chromospheric and coronal activity are commonly documented in contact binaries. The most common is the O'Connell effect manifesting as differing maximum brightness at phase 0.25 and phase 0.75. Although the O'Connell effect is modelled by the incorporation of star spots, such an approach may lead to an erroneous geometric solution even though the modelled and observed light curves appear to fit more closely \citep[See] [for a summary discussion]{2024RAA....24h5018W}. As the geometric solution forms the foundation for further analysis, we have avoided a spotted solution. There are other markers of increased chromospheric and coronal activity such as enhanced ultraviolet emissions, especially at shorter wavelengths, and enhanced x-ray emissions. Unfortunately, V717 And has not been observed at short ultraviolet wavelengths; however, it has been observed to have significantly increased x-ray emissions \citep{2022A&A...663A.115L}.

The availability of a large library of medium-resolution spectra from LAMOST has led to the investigation of longer wavelength spectroscopic indicators of enhanced chromospheric activity in contact binaries \citep{2024RAA....24h5018W, 2024ApJS..272....6H}. The LAMOST spectra, in general, demonstrate less noise in the longer infrared part and analysis is somewhat easier than in the shorter bluer part, where the continuum is not well defined \citep{2007A&A...466.1089B}. Of the three main CaII IRT lines ($\lambda$8498, $\lambda$8542, and $\lambda$8663), $\lambda$8542 and $\lambda$8663 are usually better defined, and the analysis is usually limited to these two lines. The trailing wings of the CaII IRT lines are usually quite extended and their shape is variable due to a significant impact from temperature variations in the photospheric layers \citep{2005A&A...430..669A}.  Unlike the wings, the depth of CaII IRT lines is governed by the uppermost chromospheric layers, and their central depression correlates with the degree of chromospheric activity \citep{2000A&A...353..666C}: The depth of the core relative to the continuum correlates inversely with chromospheric activity.

\citet{2024RAA....24h5018W} investigated CaII IRT filling in 10 extreme low mass contact binaries confirming enhanced chromospheric activity regardless of the presence or absence of the O'Connell effect. We follow the procedure described in \citet{2024RAA....24h5018W} to investigate the filling of the $\lambda$8542 and $\lambda$8663 lines from the LAMOST spectrum of V717 And. Briefly, we select a spectrum of a closely matching star from an atlas of template spectra \citep{2023ApJS..265...61J}. For the present study, we selected template number 513 (5903\,K, $[\mathrm{Fe}/\mathrm{H}] = -0.37$). LAMOST estimates the metallicity of V717 And to be $[\mathrm{Fe}/\mathrm{H}] = -0.38$. Due to the high rotational velocity of the contact binary, the template spectrum was broadened using the recently published script \citep{2023RNAAS...7...91C}. \citet{2007A&A...466.1089B} consider the difference in the amplitude of the central depression between the observed and template normalised spectra as an indicator of enhanced chromospheric activity. The excess filling of the CaII IRT lines of V717 And is illustrated in Figure 4. The extent of filling is comparable to that seen by \citep{2024RAA....24h5018W}.
\begin{figure}
	\includegraphics[width=\columnwidth]{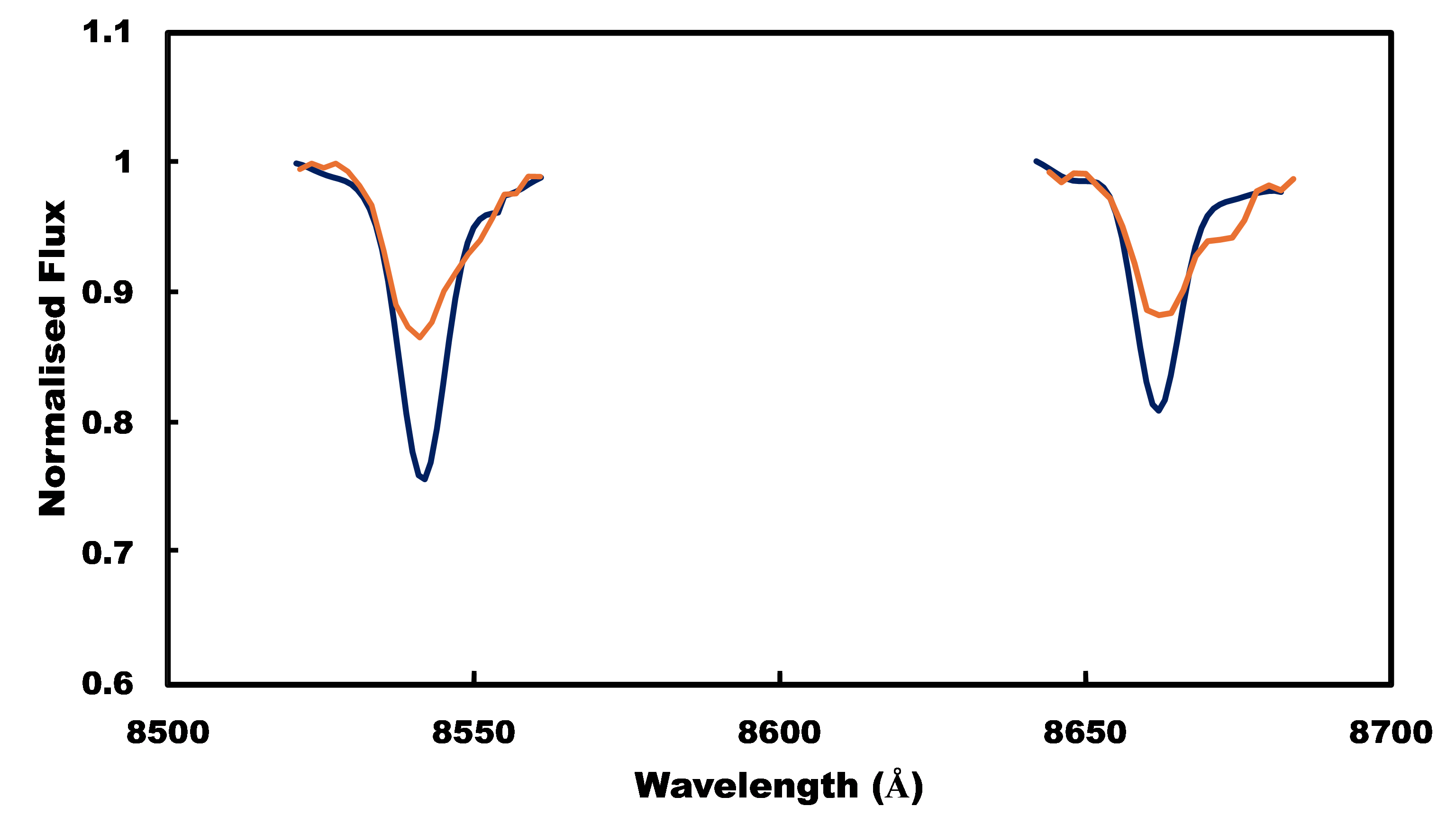}
    \caption{Filling of the two main ($\lambda$8542 and $\lambda$8663) CaII IRT lines. The orange lines represent the flux of V717 And and the purple lines represent the flux of the template star}
    \end{figure}

\subsection{Orbital Stability}

Ever since the identification of V1309 Sco as a contact binary merger event \citep{2011A&A...528A.114T}, there has been a significant increase in the identification of potential merger candidates. Theoretical considerations prior to the identification of V1309 Sco had indicated merger would likely occur when the mass ratio of the components was very small and much of the research had centered on defining a universal minimum mass ratio at which merger would take place \citep{1995ApJ...444L..41R, 2007MNRAS.377.1635A}. More recent work \citep{2021MNRAS.501..229W, 2024MNRAS.535.2494W} has indicated that there is no universal instability mass ratio and each contact binary has a unique mass ratio at which it will become unstable. The instability mass ratio, for a given system, is not fixed and changes significantly even during normal main sequence evolution. Observationally there are a few indicators which point to variable and changing mass ratio of contact binaries. \citet{2023A&A...672A.176P} indicated that the instability mass ratio may depend on the period with extremely low instability mass ratio for systems with periods greater than 0.3d while the instability mass ratio was estimated significantly higher for systems with shorter periods. The paucity of observed red novae, even in the presence of significant and ever increasing number of extreme low mass ratio contact binaries, is also supportive of systems remaining stable potentially throughout their nuclear lifespans \citep{2024MNRAS.535.2494W}. 

Without considering effects of metallicity or age \citet{2021MNRAS.501..229W} showed that the instability mass ratio range between marginal ($f=0$) and full contact ($f=1$) can be estimated from two simple quadratic relations linked to the mass of the primary component:

\begin{equation}
    q_{inst}=0.1269M_{1}^2-0.4496M_{1}+0.4403 \ (f=1)
\end{equation} 
and
\begin{equation}
  q_{inst}=0.0772M_{1}^2-0.3003M_{1}+0.3237 \ (f=0).  
\end{equation}

Based on the above relations the instability mass ratio range for V717 And would be 0.100--0.118. \citet{2024MNRAS.535.2494W} further showed that if age and metallicity are taken into account, systems with metallicity less than Solar metallicity and older systems may well be more stable. Assuming the primary is half way through its main sequence evolution and adopting the LAMOST metallicity of $[\mathrm{Fe}/\mathrm{H}] = -0.38$, the predicted instability mass ratio range is even lower at 0.065--0.070. The estimated mass ratio of V717 And at 0.197 is well above the instability mass ratio and as such we consider the system to be stable and not a potential merger candidate.

\section{Summary}

Orbital stability modelling suggests that contact binaries should merge. Theoretical considerations indicate such a merger would likely occur at extreme low mass ratios. No real time merger event has been documented, the only confirmed event, that of V1309 Sco, was only recognised in retrospect \citep{2011A&A...528A.114T}. Considerable advance has been made in defining likely instability parameters however finding systems that meet such criteria are rare \citep{2024MNRAS.535.2494W}. Although many hundreds of thousands of contact binary systems have been identified, only a small percentage are available for analysis without the need for large telescopes, sophisticated instruments and prohibitively long telescope times. Although methods have been developed to exclude likely stable systems based on survey photometry there is no definitive method to confirm potential orbital instability based purely on survey data. V717 And was selected from survey data because it demonstrated features of likely low mass ratio given the small amplitude and long total eclipse duration. 

Photometric analysis of V171 And shows that the system is an extreme low mass ratio contact binary with a steady period. The system does show signs of significant chromospheric activity with the presence of the O'Connell effect in the $B$-band where there is slight increase in the overall light curve amplitude and low grade difference in the two maxima. Our analysis and other investigations also point to significant spectral indicators of increased chromospheric activity such as CaII IRT line filling. 

Although the system is of extreme low mass ratio, orbital stability analysis suggests that the system is stable and not a potential merger candidate at this time. The presence of enhanced chromospheric activity although not a marker of orbital stability it can be a marker of enhanced magnetic activity. Enhanced magnetic activity may indicate significant magnetic stellar winds and potential enhanced angular momentum loss and period instability. The system, although not currently unstable, should at least undergo regular period monitoring.

\section*{Acknowledgements}

\noindent MG acknowledges funding provided by the Ministry of Science, Technological Development and Innovation of the Republic of Serbia through the contract 451-03-136/2025-03/200002.\\

\noindent This work has made use of data from the European Space Agency (ESA) mission
{\it Gaia} (\url{https://www.cosmos.esa.int/gaia}), processed by the {\it Gaia}
Data Processing and Analysis Consortium (DPAC,
\url{https://www.cosmos.esa.int/web/gaia/dpac/consortium}). Funding for the DPAC
has been provided by national institutions, in particular the institutions
participating in the {\it Gaia} Multilateral Agreement.\\

\bibliographystyle{mnras}
\bibliography{rgebs} 

\begin{thebibliography}{}
\makeatletter
\relax
\def\mn@urlcharsother{\let\do\@makeother \do\$\do\&\do\#\do\^\do\_\do\%\do\~}
\def\mn@doi{\begingroup\mn@urlcharsother \@ifnextchar [ {\mn@doi@} {\mn@doi@[]}}
\def\mn@doi@[#1]#2{\def\@tempa{#1}\ifx\@tempa\@empty \href {http://dx.doi.org/#2} {doi:#2}\else \href {http://dx.doi.org/#2} {#1}\fi \endgroup}
\def\mn@eprint#1#2{\mn@eprint@#1:#2::\@nil}
\def\mn@eprint@arXiv#1{\href {http://arxiv.org/abs/#1} {{\tt arXiv:#1}}}
\def\mn@eprint@dblp#1{\href {http://dblp.uni-trier.de/rec/bibtex/#1.xml} {dblp:#1}}
\def\mn@eprint@#1:#2:#3:#4\@nil{\def\@tempa {#1}\def\@tempb {#2}\def\@tempc {#3}\ifx \@tempc \@empty \let \@tempc \@tempb \let \@tempb \@tempa \fi \ifx \@tempb \@empty \def\@tempb {arXiv}\fi \@ifundefined {mn@eprint@\@tempb}{\@tempb:\@tempc}{\expandafter \expandafter \csname mn@eprint@\@tempb\endcsname \expandafter{\@tempc}}}

\bibitem[\protect\citeauthoryear{{Andretta}, {Bus{\`a}}, {Gomez}  \& {Terranegra}}{{Andretta} et~al.}{2005}]{2005A&A...430..669A}
{Andretta} V.,  {Bus{\`a}} I.,  {Gomez} M.~T.,   {Terranegra} L.,  2005, \mn@doi [\aap] {10.1051/0004-6361:20041745}, \href {https://ui.adsabs.harvard.edu/abs/2005A&A...430..669A} {430, 669}

\bibitem[\protect\citeauthoryear{{Arbutina}}{{Arbutina}}{2007}]{2007MNRAS.377.1635A}
{Arbutina} B.,  2007, \mn@doi [\mnras] {10.1111/j.1365-2966.2007.11723.x}, \href {https://ui.adsabs.harvard.edu/abs/2007MNRAS.377.1635A} {377, 1635}

\bibitem[\protect\citeauthoryear{{Awadalla} \& {Hanna}}{{Awadalla} \& {Hanna}}{2005}]{2005JKAS...38...43A}
{Awadalla} N.~S.,  {Hanna} M.~A.,  2005, \mn@doi [Journal of Korean Astronomical Society] {10.5303/JKAS.2005.38.2.043}, \href {https://ui.adsabs.harvard.edu/abs/2005JKAS...38...43A} {38, 43}

\bibitem[\protect\citeauthoryear{{Bus{\`a}}, {Aznar Cuadrado}, {Terranegra}, {Andretta}  \& {Gomez}}{{Bus{\`a}} et~al.}{2007}]{2007A&A...466.1089B}
{Bus{\`a}} I.,  {Aznar Cuadrado} R.,  {Terranegra} L.,  {Andretta} V.,   {Gomez} M.~T.,  2007, \mn@doi [\aap] {10.1051/0004-6361:20065588}, \href {https://ui.adsabs.harvard.edu/abs/2007A&A...466.1089B} {466, 1089}

\bibitem[\protect\citeauthoryear{{Carvalho} \& {Johns-Krull}}{{Carvalho} \& {Johns-Krull}}{2023}]{2023RNAAS...7...91C}
{Carvalho} A.,  {Johns-Krull} C.~M.,  2023, \mn@doi [Research Notes of the American Astronomical Society] {10.3847/2515-5172/acd37e}, \href {https://ui.adsabs.harvard.edu/abs/2023RNAAS...7...91C} {7, 91}

\bibitem[\protect\citeauthoryear{{Chen}, {Wang}, {Deng}, {de Grijs}, {Yang}  \& {Tian}}{{Chen} et~al.}{2020}]{2020ApJS..249...18C}
{Chen} X.,  {Wang} S.,  {Deng} L.,  {de Grijs} R.,  {Yang} M.,   {Tian} H.,  2020, \mn@doi [\apjs] {10.3847/1538-4365/ab9cae}, \href {https://ui.adsabs.harvard.edu/abs/2020ApJS..249...18C} {249, 18}

\bibitem[\protect\citeauthoryear{{Chmielewski}}{{Chmielewski}}{2000}]{2000A&A...353..666C}
{Chmielewski} Y.,  2000, \aap, \href {https://ui.adsabs.harvard.edu/abs/2000A&A...353..666C} {353, 666}

\bibitem[\protect\citeauthoryear{{Collins}, {Kielkopf}, {Stassun}  \& {Hessman}}{{Collins} et~al.}{2017}]{2017AJ....153...77C}
{Collins} K.~A.,  {Kielkopf} J.~F.,  {Stassun} K.~G.,   {Hessman} F.~V.,  2017, \mn@doi [\aj] {10.3847/1538-3881/153/2/77}, \href {https://ui.adsabs.harvard.edu/abs/2017AJ....153...77C} {153, 77}

\bibitem[\protect\citeauthoryear{{Devarapalli}, {Jagirdar}, {Prasad}, {Thomas}, {Ahmed}, {Gralapally}  \& {Das}}{{Devarapalli} et~al.}{2020}]{2020MNRAS.493.1565D}
{Devarapalli} S.~P.,  {Jagirdar} R.,  {Prasad} R.~M.,  {Thomas} V.~S.,  {Ahmed} S.~A.,  {Gralapally} R.,   {Das} J.~P.,  2020, \mn@doi [\mnras] {10.1093/mnras/staa031}, \href {https://ui.adsabs.harvard.edu/abs/2020MNRAS.493.1565D} {493, 1565}

\bibitem[\protect\citeauthoryear{{Drake} et~al.,}{{Drake} et~al.}{2014}]{2014ApJS..213....9D}
{Drake} A.~J.,  et~al., 2014, \mn@doi [\apjs] {10.1088/0067-0049/213/1/9}, \href {https://ui.adsabs.harvard.edu/abs/2014ApJS..213....9D} {213, 9}

\bibitem[\protect\citeauthoryear{{Gaia Collaboration} et~al.,}{{Gaia Collaboration} et~al.}{2016}]{2016A&A...595A...1G}
{Gaia Collaboration} et~al., 2016, \mn@doi [\aap] {10.1051/0004-6361/201629272}, \href {https://ui.adsabs.harvard.edu/abs/2016A&A...595A...1G} {595, A1}

\bibitem[\protect\citeauthoryear{{Gaia Collaboration} et~al.,}{{Gaia Collaboration} et~al.}{2023}]{2023A&A...674A...1G}
{Gaia Collaboration} et~al., 2023, \mn@doi [\aap] {10.1051/0004-6361/202243940}, \href {https://ui.adsabs.harvard.edu/abs/2023A&A...674A...1G} {674, A1}

\bibitem[\protect\citeauthoryear{{Henden}, {Levine}, {Terrell}  \& {Welch}}{{Henden} et~al.}{2015}]{2015AAS...22533616H}
{Henden} A.~A.,  {Levine} S.,  {Terrell} D.,   {Welch} D.~L.,  2015, in American Astronomical Society Meeting Abstracts \#225. p. 336.16

\bibitem[\protect\citeauthoryear{{Huang} et~al.,}{{Huang} et~al.}{2024}]{2024ApJS..272....6H}
{Huang} X.,  et~al., 2024, \mn@doi [\apjs] {10.3847/1538-4365/ad33bc}, \href {https://ui.adsabs.harvard.edu/abs/2024ApJS..272....6H} {272, 6}

\bibitem[\protect\citeauthoryear{{Jayasinghe} et~al.,}{{Jayasinghe} et~al.}{2018}]{2018MNRAS.477.3145J}
{Jayasinghe} T.,  et~al., 2018, \mn@doi [\mnras] {10.1093/mnras/sty838}, \href {https://ui.adsabs.harvard.edu/abs/2018MNRAS.477.3145J} {477, 3145}

\bibitem[\protect\citeauthoryear{{Ji}, {Liu}, {Deng}, {Zhang}, {Li}, {Tian}  \& {Li}}{{Ji} et~al.}{2023}]{2023ApJS..265...61J}
{Ji} W.,  {Liu} C.,  {Deng} L.,  {Zhang} B.,  {Li} J.,  {Tian} H.,   {Li} J.,  2023, \mn@doi [\apjs] {10.3847/1538-4365/acbf42}, \href {https://ui.adsabs.harvard.edu/abs/2023ApJS..265...61J} {265, 61}

\bibitem[\protect\citeauthoryear{{Li}, {Han}  \& {Zhang}}{{Li} et~al.}{2004}]{2004MNRAS.355.1383L}
{Li} L.,  {Han} Z.,   {Zhang} F.,  2004, \mn@doi [\mnras] {10.1111/j.1365-2966.2004.08457.x}, \href {https://ui.adsabs.harvard.edu/abs/2004MNRAS.355.1383L} {355, 1383}

\bibitem[\protect\citeauthoryear{{Li}, {Zhu}, {Ding}, {Xu}, {Zheng}, {Qiu}  \& {Liu}}{{Li} et~al.}{2024}]{2024ApJS..271...32L}
{Li} X.-Z.,  {Zhu} Q.-F.,  {Ding} X.,  {Xu} X.-H.,  {Zheng} H.,  {Qiu} J.-S.,   {Liu} M.-C.,  2024, \mn@doi [\apjs] {10.3847/1538-4365/ad226a}, \href {https://ui.adsabs.harvard.edu/abs/2024ApJS..271...32L} {271, 32}

\bibitem[\protect\citeauthoryear{{Liu}, {Wu}, {Esamdin}, {Gu}, {Sun}  \& {Wang}}{{Liu} et~al.}{2022}]{2022A&A...663A.115L}
{Liu} J.,  {Wu} J.,  {Esamdin} A.,  {Gu} W.-M.,  {Sun} M.,   {Wang} J.,  2022, \mn@doi [\aap] {10.1051/0004-6361/202142963}, \href {https://ui.adsabs.harvard.edu/abs/2022A&A...663A.115L} {663, A115}

\bibitem[\protect\citeauthoryear{{Luo}, {Zhao}, {Zhao}  \& {et al.}}{{Luo} et~al.}{2022}]{2022yCat.5156....0L}
{Luo} A.~L.,  {Zhao} Y.~H.,  {Zhao} G.,   {et al.} 2022, {VizieR Online Data Catalog: LAMOST DR7 catalogs (Luo+, 2019)}, VizieR On-line Data Catalog: V/156. Originally published in: 2019RAA..in.prep..L

\bibitem[\protect\citeauthoryear{{Mochnacki}}{{Mochnacki}}{1981}]{1981ApJ...245..650M}
{Mochnacki} S.~W.,  1981, \mn@doi [\apj] {10.1086/158841}, \href {https://ui.adsabs.harvard.edu/abs/1981ApJ...245..650M} {245, 650}

\bibitem[\protect\citeauthoryear{{Nelson}}{{Nelson}}{2021}]{2021NewA...8601565N}
{Nelson} R.~H.,  2021, \mn@doi [\na] {10.1016/j.newast.2020.101565}, \href {https://ui.adsabs.harvard.edu/abs/2021NewA...8601565N} {86, 101565}

\bibitem[\protect\citeauthoryear{{Paunzen} \& {Vanmunster}}{{Paunzen} \& {Vanmunster}}{2016}]{2016AN....337..239P}
{Paunzen} E.,  {Vanmunster} T.,  2016, \mn@doi [Astronomische Nachrichten] {10.1002/asna.201512254}, \href {https://ui.adsabs.harvard.edu/abs/2016AN....337..239P} {337, 239}

\bibitem[\protect\citeauthoryear{{Pecaut} \& {Mamajek}}{{Pecaut} \& {Mamajek}}{2013}]{2013ApJS..208....9P}
{Pecaut} M.~J.,  {Mamajek} E.~E.,  2013, \mn@doi [\apjs] {10.1088/0067-0049/208/1/9}, \href {https://ui.adsabs.harvard.edu/abs/2013ApJS..208....9P} {208, 9}

\bibitem[\protect\citeauthoryear{{Pe{\v{s}}ta} \& {Pejcha}}{{Pe{\v{s}}ta} \& {Pejcha}}{2023}]{2023A&A...672A.176P}
{Pe{\v{s}}ta} M.,  {Pejcha} O.,  2023, \mn@doi [\aap] {10.1051/0004-6361/202245613}, \href {https://ui.adsabs.harvard.edu/abs/2023A&A...672A.176P} {672, A176}

\bibitem[\protect\citeauthoryear{{Rasio}}{{Rasio}}{1995}]{1995ApJ...444L..41R}
{Rasio} F.~A.,  1995, \mn@doi [\apjl] {10.1086/187855}, \href {https://ui.adsabs.harvard.edu/abs/1995ApJ...444L..41R} {444, L41}

\bibitem[\protect\citeauthoryear{{Robertson} \& {Eggleton}}{{Robertson} \& {Eggleton}}{1977}]{1977MNRAS.179..359R}
{Robertson} J.~A.,  {Eggleton} P.~P.,  1977, \mn@doi [\mnras] {10.1093/mnras/179.3.359}, \href {https://ui.adsabs.harvard.edu/abs/1977MNRAS.179..359R} {179, 359}

\bibitem[\protect\citeauthoryear{{Samec}, {Labadorf}, {Hawkins}, {Faulkner}  \& {Van Hamme}}{{Samec} et~al.}{2011}]{2011AJ....142..117S}
{Samec} R.~G.,  {Labadorf} C.~M.,  {Hawkins} N.~C.,  {Faulkner} D.~R.,   {Van Hamme} W.,  2011, \mn@doi [\aj] {10.1088/0004-6256/142/4/117}, \href {https://ui.adsabs.harvard.edu/abs/2011AJ....142..117S} {142, 117}

\bibitem[\protect\citeauthoryear{{Schwarzenberg-Czerny}}{{Schwarzenberg-Czerny}}{1996}]{1996ApJ...460L.107S}
{Schwarzenberg-Czerny} A.,  1996, \mn@doi [\apjl] {10.1086/309985}, \href {https://ui.adsabs.harvard.edu/abs/1996ApJ...460L.107S} {460, L107}

\bibitem[\protect\citeauthoryear{{Sergey}}{{Sergey}}{2013}]{2013PZP....13....9S}
{Sergey} I.,  2013, Peremennye Zvezdy Prilozhenie, \href {https://ui.adsabs.harvard.edu/abs/2013PZP....13....9S} {13, 9}

\bibitem[\protect\citeauthoryear{{Shappee} et~al.,}{{Shappee} et~al.}{2014}]{2014ApJ...788...48S}
{Shappee} B.~J.,  et~al., 2014, \mn@doi [\apj] {10.1088/0004-637X/788/1/48}, \href {https://ui.adsabs.harvard.edu/abs/2014ApJ...788...48S} {788, 48}

\bibitem[\protect\citeauthoryear{{Terrell} \& {Wilson}}{{Terrell} \& {Wilson}}{2005}]{2005Ap&SS.296..221T}
{Terrell} D.,  {Wilson} R.~E.,  2005, \mn@doi [\apss] {10.1007/s10509-005-4449-4}, \href {https://ui.adsabs.harvard.edu/abs/2005Ap&SS.296..221T} {296, 221}

\bibitem[\protect\citeauthoryear{{Thiemann}, {Norton}, {Dickinson}, {McMaster}  \& {Kolb}}{{Thiemann} et~al.}{2021}]{2021MNRAS.502.1299T}
{Thiemann} H.~B.,  {Norton} A.~J.,  {Dickinson} H.~J.,  {McMaster} A.,   {Kolb} U.~C.,  2021, \mn@doi [\mnras] {10.1093/mnras/stab140}, \href {https://ui.adsabs.harvard.edu/abs/2021MNRAS.502.1299T} {502, 1299}

\bibitem[\protect\citeauthoryear{{Tylenda} et~al.,}{{Tylenda} et~al.}{2011}]{2011A&A...528A.114T}
{Tylenda} R.,  et~al., 2011, \mn@doi [\aap] {10.1051/0004-6361/201016221}, \href {https://ui.adsabs.harvard.edu/abs/2011A&A...528A.114T} {528, A114}

\bibitem[\protect\citeauthoryear{{Vilhu}}{{Vilhu}}{1981}]{1981Ap&SS..78..401V}
{Vilhu} O.,  1981, \mn@doi [\apss] {10.1007/BF00648946}, \href {https://ui.adsabs.harvard.edu/abs/1981Ap&SS..78..401V} {78, 401}

\bibitem[\protect\citeauthoryear{{Wadhwa}, {De Horta}, {Filipovi{\'c}}, {Tothill}, {Arbutina}, {Petrovi{\'c}}  \& {Djura{\v{s}}evi{\'c}}}{{Wadhwa} et~al.}{2021}]{2021MNRAS.501..229W}
{Wadhwa} S.~S.,  {De Horta} A.,  {Filipovi{\'c}} M.~D.,  {Tothill} N.~F.~H.,  {Arbutina} B.,  {Petrovi{\'c}} J.,   {Djura{\v{s}}evi{\'c}} G.,  2021, \mn@doi [\mnras] {10.1093/mnras/staa3637}, \href {https://ui.adsabs.harvard.edu/abs/2021MNRAS.501..229W} {501, 229}

\bibitem[\protect\citeauthoryear{{Wadhwa}, {de Horta}, {Filipovi{\'c}}  \& {Totohill}}{{Wadhwa} et~al.}{2022a}]{2022JApA...43...42W}
{Wadhwa} S.~S.,  {de Horta} A.~Y.,  {Filipovi{\'c}} M.~D.,   {Totohill} F.~H.~N.,  2022a, \mn@doi [Journal of Astrophysics and Astronomy] {10.1007/s12036-022-09832-9}, \href {https://ui.adsabs.harvard.edu/abs/2022JApA...43...42W} {43, 42}

\bibitem[\protect\citeauthoryear{{Wadhwa}, {De Horta}, {Filipovi{\'c}}, {Tothill}, {Arbutina}, {Petrovi{\'c}}  \& {Djura{\v{s}}evi{\'c}}}{{Wadhwa} et~al.}{2022b}]{2022JApA...43...94W}
{Wadhwa} S.~S.,  {De Horta} A.~Y.,  {Filipovi{\'c}} M.~D.,  {Tothill} N. F.~H.,  {Arbutina} B.,  {Petrovi{\'c}} J.,   {Djura{\v{s}}evi{\'c}} G.,  2022b, \mn@doi [Journal of Astrophysics and Astronomy] {10.1007/s12036-022-09888-7}, \href {https://ui.adsabs.harvard.edu/abs/2022JApA...43...94W} {43, 94}

\bibitem[\protect\citeauthoryear{{Wadhwa}, {Petrovi{\'c}}, {Tothill}, {De Horta}, {Filipovi{\'c}}  \& {Djura{\v{s}}evi{\'c}}}{{Wadhwa} et~al.}{2023}]{2023RAA....23k5001W}
{Wadhwa} S.~S.,  {Petrovi{\'c}} J.,  {Tothill} N. F.~H.,  {De Horta} A.~Y.,  {Filipovi{\'c}} M.~D.,   {Djura{\v{s}}evi{\'c}} G.,  2023, \mn@doi [Research in Astronomy and Astrophysics] {10.1088/1674-4527/acf445}, \href {https://ui.adsabs.harvard.edu/abs/2023RAA....23k5001W} {23, 115001}

\bibitem[\protect\citeauthoryear{{Wadhwa}, {Popowicz}, {Michel}, {Kosti{\'c}}, {Vince}, {Tothill}, {De Horta}  \& {Filipovi{\'c}}}{{Wadhwa} et~al.}{2024a}]{2024RAA....24h5018W}
{Wadhwa} S.~S.,  {Popowicz} A.,  {Michel} R.,  {Kosti{\'c}} P.,  {Vince} O.,  {Tothill} N. F.~H.,  {De Horta} A.~Y.,   {Filipovi{\'c}} M.~D.,  2024a, \mn@doi [Research in Astronomy and Astrophysics] {10.1088/1674-4527/ad621f}, \href {https://ui.adsabs.harvard.edu/abs/2024RAA....24h5018W} {24, 085018}

\bibitem[\protect\citeauthoryear{{Wadhwa}, {Landin}, {Arbutina}, {Tothill}, {De Horta}, {Filipovi{\'c}}, {Petrovi{\'c}}  \& {Djura{\v{s}}evi{\'c}}}{{Wadhwa} et~al.}{2024b}]{2024MNRAS.535.2494W}
{Wadhwa} S.~S.,  {Landin} N.~R.,  {Arbutina} B.,  {Tothill} N. F.~H.,  {De Horta} A.~Y.,  {Filipovi{\'c}} M.~D.,  {Petrovi{\'c}} J.,   {Djura{\v{s}}evi{\'c}} G.,  2024b, \mn@doi [\mnras] {10.1093/mnras/stae2511}, \href {https://ui.adsabs.harvard.edu/abs/2024MNRAS.535.2494W} {535, 2494}

\bibitem[\protect\citeauthoryear{{Wilson}}{{Wilson}}{1990}]{1990ApJ...356..613W}
{Wilson} R.~E.,  1990, \mn@doi [\apj] {10.1086/168867}, \href {https://ui.adsabs.harvard.edu/abs/1990ApJ...356..613W} {356, 613}

\bibitem[\protect\citeauthoryear{{Wilson} \& {Devinney}}{{Wilson} \& {Devinney}}{1971}]{1971ApJ...166..605W}
{Wilson} R.~E.,  {Devinney} E.~J.,  1971, \mn@doi [\apj] {10.1086/150986}, \href {https://ui.adsabs.harvard.edu/abs/1971ApJ...166..605W} {166, 605}

\bibitem[\protect\citeauthoryear{{Xu}, {Zhu}, {Thawicharat}  \& {Boonrucksar}}{{Xu} et~al.}{2022}]{2022RAA....22c5024X}
{Xu} H.-S.,  {Zhu} L.-Y.,  {Thawicharat} S.,   {Boonrucksar} S.,  2022, \mn@doi [Research in Astronomy and Astrophysics] {10.1088/1674-4527/ac4ca5}, \href {https://ui.adsabs.harvard.edu/abs/2022RAA....22c5024X} {22, 035024}

\bibitem[\protect\citeauthoryear{{Yildiz} \& {Do{\u{g}}an}}{{Yildiz} \& {Do{\u{g}}an}}{2013}]{2013MNRAS.430.2029Y}
{Yildiz} M.,  {Do{\u{g}}an} T.,  2013, \mn@doi [\mnras] {10.1093/mnras/stt028}, \href {https://ui.adsabs.harvard.edu/abs/2013MNRAS.430.2029Y} {430, 2029}

\bibitem[\protect\citeauthoryear{{Zhang}, {Zhang}, {He}, {Luo}  \& {Zhang}}{{Zhang} et~al.}{2024}]{2024A&A...688A..23Z}
{Zhang} W.,  {Zhang} J.,  {He} H.,  {Luo} A.,   {Zhang} H.,  2024, \mn@doi [\aap] {10.1051/0004-6361/202348988}, \href {https://ui.adsabs.harvard.edu/abs/2024A&A...688A..23Z} {688, A23}

\bibitem[\protect\citeauthoryear{{van Hamme}}{{van Hamme}}{1993}]{1993AJ....106.2096V}
{van Hamme} W.,  1993, \mn@doi [\aj] {10.1086/116788}, \href {https://ui.adsabs.harvard.edu/abs/1993AJ....106.2096V} {106, 2096}

\makeatother
\end{thebibliography}


\end{document}